\documentclass[journal=jpcafh,manuscript=article]{achemso}

\usepackage[version=3]{mhchem} 
\usepackage{amsmath,amssymb,amsfonts}
\usepackage{xcolor} 
\usepackage{epsfig}
\usepackage{amssymb}
\usepackage{subfig}
\usepackage{graphicx}
\usepackage{epsfig}
\usepackage{longtable}
\usepackage{bm}
\usepackage{array}
\usepackage{lscape}
\usepackage{amsmath}

\usepackage{subfig}
\usepackage{subfloat}

\author{Aleksander P. Wo{\'z}niak}
\affiliation[University of Warsaw]{Faculty of Chemistry, University of Warsaw, Pasteura 1, 02-093 Warsaw, Poland}
\email{ap.wozniak@uw.edu.pl}
\author{Ludwik Adamowicz}
\affiliation[University of Arizona]{Department of Chemistry and Biochemistry, University of Arizona,  
1306 E University Blvd,
Tucson, Arizona 85721-0041, USA}
\author{Thomas Bondo Pedersen}
\author{Simen Kvaal}
\affiliation[University of Oslo]{Hylleraas Centre for Quantum Molecular Sciences, Department of Chemistry, University of Oslo, 
P.O. Box 1033 Blindern, N-0315 Oslo, Norway}

\title[Gaussians for Quantum Dynamics]{Rothe Time Propagation for Coupled Electronic and Rovibrational Quantum Dynamics}

\begin{document}


\begin{abstract}    
When time-propagating a wave packet representing a molecular
system interacting with strong attosecond laser pulses, one needs to
use an approach that is capable of describing intricate
coupled electronic-nuclear events that require departure from the
conventional adiabatic Born-Oppenheimer (BO) approximation.
Hence, the propagation should be carried out simultaneously
for the electrons and nuclei, treating both particle types on an equal footing
\emph{without} invoking the BO approximation. 
In such calculations, in order to achieve high accuracy, the wave packet 
needs to be expanded in basis functions that explicitly depend on
interparticle distances, such as all-particle explicitly correlated Gaussians
(ECGs).
In our previous work,
we employed basis sets consisting of ECGs with optimizable 
complex exponential parameters to fit time-dependent wave functions obtained from grid-based 
propagations of two model systems: a nucleus in a Morse potential 
and an electron in a central-field Coulomb-like potential, subjected to intense laser pulses.
In this work, we present a proof-of-principle study of the time propagation of
linear combinations of ECGs for these two models using Rothe's method. 
It is shown that the approach very closely reproduce the virtually exact results of grid-based 
propagation for both systems. This provides further evidence that ECGs constitute a viable alternative to purely grid-based
simulations of coupled nuclear-electronic dynamics driven by intense laser pulses.

\end{abstract}

\section{Introduction} \label{introduction}

Experimental studies of the dynamics of chemical and physical processes
at the atomistic level usually involve simulations to aid interpretation.
This is particularly true for events
triggered by the interaction of individual atoms, molecules,
or simple clusters with high-energy photons.
New experimental techniques for manipulating atoms and
molecules with intense ultrashort (atto- and few-femtosecond)
laser pulses has revealed the complicated nature of the dynamics
of these experiments. To ensure correct interpretation of the
experimental observations, reliable quantum-dynamics (QD)
simulations of the experiments need to be performed.
As the intense broad-band laser pulses deliver high energy to a molecule,
all degrees of freedom are affected. These include motions associated
with rovibrational transitions and motions 
associated with electronic and vibronic transitions, as
well as motions resulting in ionization and/or fragmentation 
of the system.
In all these motions the coupling of the electronic and nuclear degrees
of freedom plays a significant role.
Thus, QD simulations of the molecular systems interacting with
ultrashort intense laser pulses
carried out by integrating the time-dependent 
Schr\"{o}dinger equation (TDSE), have to include a proper
description of all these effects.
In particular, the approach used has to depart from the
Born-Oppenheimer (BO) approximation \cite{Born1927,Born1954}
because only then can the coupling of the electronic and nuclear
degrees of freedom be accurately described.
This is the approach investigated in the present work.

If a finite basis set is
employed in the simulation, the basis functions that are used to
expand the time-evolving wave packet representing the system
must be capable of describing an array of effects.
These include the correlation effects associated with 
the Coulombic electron-electron, electron-nucleus, and 
nucleus-nucleus interaction of which the latter is the
strongest. It happens because heavy nuclei with alike charges
stay very strictly separated in their motions. This is different
for the much lighter electrons. Their wave functions are wider and overlap,
resulting in a noticeable probability of finding two electrons in the
same point in space. The electronic-nuclear correlation is also strong, as the
electron, particularly the core electrons, follow the nuclei
in their motion very closely.

The other effects that the basis functions used in the QD simulation
need to describe are the oscillations and axial deformation 
of the wave functions along the field direction,
caused by rotation of the direction and oscillation of
the intensity of the laser pulse.
The oscillations can also be caused by the wave packet 
acquiring contributions from states representing a 
large range of the rovibrational and electronic 
excitations of the system. 

All the above mentioned effects necessitate the use of very 
flexible basis functions, whose size must be adjusted in the simulation
process and whose parameters (linear and non-linear) are 
thoroughly optimized at every simulation step.
The basis functions used in this work and the procedure used
in the laser-induced time propagation of the wave packet 
representing the system are described in the next section.
To best describe the mentioned effects,
particularly the coupling effects, using a basis-set approach,
one needs to expand the wave function in terms 
of functions that explicitly depend on
the interparticle coordinates, i.e. 
nucleus-nucleus, nucleus-electron, and 
electron-electron distances. Such functions are called 
explicitly-correlated functions (ECFs). 
In the first part of this work, we 
review the ECFs used in the atomic and molecular
calculations of stationary bound states and
we discuss the features of these functions
that make them particularly useful in
QM calculations of atomic and molecular systems.
In particular, we focus on all-particle Gaussian ECFs---explicitly correlated Gaussians (ECGs)---as
these are the most popular functions 
used in non-BO high-accuracy atomic and molecular 
calculations\cite{Bubin2013,Kozlowski19932007,Mitroy2013,Bubin2005377,Matyus2019590,Simmen2014,Muolo2018_JCP,Muolo2018,Matyus2012,Muolo2020,Johnson2016,Valeev2006,Ten-no2003152,Takatsuka2003859,Varga2019}.
Moreover, shifted ECGs have been shown to correctly capture impulsive laser alignment without the BO approximation\cite{Adamowicz2022}

The main focus of the present work is the effectiveness of
ECGs. The Gaussians used in the present work combine two features
that have been proven instrumental in describing the topology
of the wave function of very highly 
excited rovibrational states of diatomic molecules
and the deformation of the highly correlated wave function 
of a molecule interacting with a static electric field\cite{Cafiero2003113,Bubin2005377,Cafiero2002033002/1,Cafiero2002073001/1}.
The mentioned features include the shifts of the centers of the ECGs
away from the origin of the internal coordinate system and making
both the elements of the matrix of the exponential
parameters of the Gaussians and the elements of the
vector of the shifts to be complex numbers. These extensions allow us to describe the radial
and angular oscillations of the wave function. It also allows us to 
describe the interparticle correlation, in particular the strong
internuclear correlation, as it was shown previously for a diatomic system\cite{Bubin2020204102,Chavez2019147,Bubin2017b,Bubin2016122}.
The complex shift coordinates together with the
complex exponential parameters enable one to describe dissociation
and fragmentation processes. In the present work, we do not use complex shifts, but instead an equivalent parameterization in terms of real shifts and plane-wave factors (see the next section).

Single-particle Gaussians have been extensively used as basis functions 
in both electronic-structure theory \cite{MEST} and
vibrational dynamics \cite{Heller1975,Heller1976,Heller1981}.
In electronic-structure theory the Gaussians are real-valued functions centered 
at the atomic nuclei and contracted to form atomic orbitals which, in turn, 
form a non-orthogonal basis for the expansion of molecular orbitals---see,
e.g., Ref.~\citenum{MEST} for a detailed account.

Single-particle Gaussians have been used before to study
molecular vibrational dynamics  \cite{Heller1975,Heller1976,Heller1981}
multi-electron dynamics \cite{Goings2018,Li2020,ofstad_time-dependent_2023}.
In the vibrational-dynamics approach proposed by \citeauthor{Heller1975},  
the use of complex-valued Gaussians, was motivated by these functions
being exact solutions for harmonic potentials. In this approach,
the Gaussian parameters are time-dependent variational
parameters. However, when the potential becomes anharmonic,
the time-dependent equations of motion quickly become ill-behaved and
one needs to resort to locally-harmonic approximations \citenum{Vanicek2020}.
One should also mention the use of 
single-particle complex-valued Gaussians in the
multiconfigurational time-dependent Hartree (MCTDH)
method\cite{Meyer_etal2009,worth_full_2003,worth_novel_2004,Burghardt_etal2008}. 
However, also in this case, 
ill-conditioned equations arise that hamper the use of the method.

Restricting the Gaussians to real-valued functions does not 
allow to describe such important highly nonlinear phenomena as ionization 
and high harmonic generation processes. Such processes are crucial 
in the dynamics involving short and extreme laser pulses. Some attempts 
have been made to alleviate 
this problem by augmenting the ECGs with other functions able to represent
continua have been made by \citeauthor{Coccia2022} \cite{aaronson_np_5_707_2009}, but they have not led
to many applications of the approach.

To test the ability of ECGs to describe molecular QD, 
in our previous work \cite{Wozniak20243659},
we used the virtually exact numerical wave packets determined on a 
grid to fit linear combinations of ECGs
The fitting was done for two two-dimensional (2D) models
representing laser-induced single-particle dynamics occurring in a Coulomb potential
and in a Morse potential. These two dynamics are essential elements
of the dynamics that occur when a diatomic molecule is interacting
with laser pulses, in particular, with large-frequency and high-intensity
ultra-short laser pulses. The tests were relevant to laser-induced dynamics 
of attosecond atomic and molecular events involved in attosecond experiments\cite{Corkum2007,Nisoli2017,Borrego-Varillas2022}.
In the present work, the testing of the ability of the ECGs to represent
time-evolving wave packets in the two previously considered 2D models is extended to generating these packets by solving 
the time-dependent Schr\"{o}dinger equation using a novel apprach called the Rothe time-propagation
method \cite{rothe1930,bornemann1990,horenko2004,schraderTimeEvolutionOptimization2024,schraderMultidimensionalQuantumDynamics2025}, in which discretization is done \emph{first in time}, leading to each time step being an optimization problem for finding the correct wavefunction in terms of the nonlinear parameters of the ECGs, as well as the linear expansion coefficients. The Rothe time-propagation method has been studied previously on the high-harmonic generation from a hydrogen atom and on dynamics in the Henon--Heiles potential\cite{schraderTimeEvolutionOptimization2024,schraderMultidimensionalQuantumDynamics2025}, showing encouraging results. The present study is complementary, focusing on 2D model systems in preparation for future studies of multiparticle systems without the non-Born--Oppenheimer approximation.

The presentation of this work starts with a description of the Hamiltonians
used in the calculations. Next, the Rothe time-propagation method is briefly outlined.
This is followed by a section on the ECGs used in the calculations
and a section describing the results of the test calculations.
Lastly, the main results of this work are summarized. 

\section{Theory and methods} \label{methods}

\subsection{Non-BO Hamiltonian of the molecular system}

The aim of this work is to develop a procedure for solving the 
time-dependent Schr\"{o}dinger equation (TDSE), 
${\rm i} \dot{\psi}(t) - \hat{H}(t)\psi(t) = 0$,
for a molecular system interacting with a short but
intense laser pulse.
In the procedure, the nuclei and electrons are treated on an equal footing
i.e. the BO approximation is not assumed.
The total non-relativistic all-particle non-BO molecular Hamiltonian
describing the interaction of a neutral molecule 
with an electric field with time dependent strength 
oriented along a chosen direction  
(we choose the direction along the $x$-axis of a laboratory-fixed coordinate frame)
can, in the dipole approximation, be rigorously separated into a center-of-mass (COM)
kinetic-energy operator and the internal Hamiltonian
\cite{Thomas1970,Bubin2013,Mitroy2013},
$\hat{H}(t)$. The separation is performed by 
transforming the total lab-frame Hamiltonian from
the Cartesian laboratory coordinates,
$\mathbf{R}_i$, $i = 1, \dots, N$, where $N$ is the total
number of particles in the molecule,
to a new Cartesian coordinate
system, whose the first three coordinates are the 
lab-frame coordinates of the COM and the remaining coordinates 
are internal coordinates.
The internal coordinate axes are parallel to the lab-frame axes
and have their origins located at a selected particle 
called the reference particle (particle no. $1$; it is usually the heaviest
nucleus in the molecule). The internal coordinates, $\{\boldsymbol{r}_i\}$,
are:
$\boldsymbol{r}_i = \boldsymbol{R}_{i+1}-\boldsymbol{R}_1$ where 
$i=1,\dots,n$ and $n=N-1$.
The internal Hamiltonian only depends on the internal coordinates
and have the following form in atomic units \cite{Bubin2013}:
\begin{equation}
\label{internal}
    \hat{H}(t) = \sum_{i=1}^n \left(
                       -\frac{1}{2\mu_i} \nabla_{{\bf r}_i}^2 + \frac{q_0 q_i}{r_i}
                 \right)
               + \sum_{i<j}^n \left(
                       \frac{q_i q_j}{r_{ij}} + \frac{1}{M_1} \nabla_{{\bf r}_i}^\prime \nabla _{{\bf r}_j}
                 \right)
               - \mathcal{F}(t) \sum_{i=1}^{n} q_i x_i,
\end{equation}
where $\mathcal{F}(t)$ is the time-dependent electric-field strength,
$M_1$ is the mass of the reference particle (particle $1$),
$q_i = Q_{i+1}$ ($i=0,\dots,n$), 
$\mu_i = M_1 M_{i+1}/(M_1 + M_{i+1})$ 
is the reduced mass of particle $i$ ($i = 1, \dots, n$), 
$Q_i$ and $M_i$ are the charge and mass
of particle $i$, respectively ($i = 1, \dots, N$),
$r_{ij}=| {\bf r}_i-{\bf r}_j| =
| {\bf R}_{i+1}-{\bf R}_{j+1}|$, 
$ r_i = | {\bf r}_i |$, and the prime denotes vector/matrix
transposition.
As one notices, internal Hamiltonian \eqref{internal} represents $n$
interacting particles with masses equal to the reduced masses
moving in the central Coulomb potential of the reference particle.
We refer to the particles, whose total internal energy is
described by the internal Hamiltonian, \eqref{internal},
as \emph{pseudo-particles} because, while they have the same charges
as the original particles, their
masses are changed to the reduced masses. Moreover, the permutational symmetry of the wavefunction is changed if the central nucleus is identical to some other nucleus in the system.
For $\mathcal{F}(t)=0$, the Hamiltonian \eqref{internal}
is fully symmetric (isotropic or atom-like) 
with respect to all rotations around the center of the
internal coordinate system. Thus, the eigenfunctions of \eqref{internal}
transform as irreducible representations 
of the fully symmetric group of rotations.
However, when $\mathcal{F}(t) \neq 0$, 
the symmetry of \eqref{internal}
is reduced to cylindrical with the symmetry axis being 
the field direction (here the $x$-axis).

\subsection{Grid-based calculations of 2D model systems}

In the internal Hamiltonian (\ref{internal}) for a diatomic system, 
the potential acting on the second nucleus that results 
from the interaction of this nucleus with the charge of the 
reference nucleus and the electrons can effectively be represented 
by a Morse-like potential. An important effect that also 
determines the electronic-nuclear dynamics of the system 
is the electrostatic attractive interaction of each of 
the electrons with the reference nucleus located 
at the center of the internal coordinate system. 
Thus, the dynamics of an electron interacting with a central 
potential created by the charge of the reference nucleus
and the dynamics of a particle interacting with a Morse potential 
are central to understanding the laser-induced dynamics of a diatomic molecule. 
The dynamics of these two 
models is investigated in 2D in the present work.

As a near-exact numerical reference for the Rothe propagations 
performed in this work, we utilize the results of grid-based simulations of the Morse model 
and of the Coulomb model presented in our previous paper, Ref. {\citenum{Wozniak20243659}}.
The detailed methodology and the justification for the choice of the simulation parameters are 
provided there, while here we only summarize the key assumptions of the calculations for the sake of 
completeness.
For both 2D models we use a Hamiltonian of the form:
\begin{equation}
    \hat{H}(t) = -\frac{1}{2\mu}
    \left (\frac{\partial^2}{\partial x^2}
    +      \frac{\partial^2}{\partial y^2} \right ) 
    + V(x,y) - q x \mathcal{F}(t), 
\end{equation}
where for the electron we use the soft Coulomb potential,
$V(x,y) = -(x^2 + y^2 + 1/2)^{-1/2}$, which 
mimics the nuclear potential of a hydrogen-like atom, and
for the Morse potential we use
$V(x,y) = D_e 
\left[ 1 - \exp (-\alpha ((x^2+y^2)^{1/2} - r_e)) \right]^2$, with $D_e = 0.17449$, $r_e = 1.4011$, 
and $\alpha = 1.4556$.
The charge and (reduced) mass are set to $q=-1$, $\mu=1$ for the
Coulomb model, and $q=1$, $\mu=1605.587$ for the Morse model.
The electric-field strength
is nonzero only in the time 
interval $t_0 < t < t_1$, where it is equal to:
\begin{equation}
    \mathcal{F}(t) = \mathcal{E}_0\sin^2 \left(\pi \frac{t-t_0}{t_1 - t_0}\right)\cos(\omega (t-\bar{t})), \qquad \bar{t} = \frac{t_0 + t_1}{2},
\end{equation}
where $\mathcal{E}_0$ denotes the maximum electric field amplitude. 
In our calculations for both models we set $t_0 = 0$.
For the Coulomb model we set $\omega = 0.25$ a.u., $t_1 
= 60$ a.u. and $\mathcal{E}_0 = 0.4$ a.u., which 
corresponds to a laser pulse of wavelength $\approx 182$ nm, with a duration of 2.5 optical cycles (foot-to-foot duration $1.45\,\text{fs}$) and a peak intensity of $5.6 \times 10^{15}\,\text{W/cm}^2$.
For the Morse model we set $\omega = 0.0$ a.u., $t_1 
= 20$ a.u. ($0.48\,\text{fs}$) and $\mathcal{E}_0 = 2.0$ a.u., 
corresponding to a short, delta-like pulse which violently but very briefly pushes the particle in the positive $x$-direction. 
We consider the dynamics for times $0 \leq t \leq 100$ a.u. for the Coulomb model and $0 \leq t \leq 
300$ a.u.~for the Morse model, including periods of free evolution after the laser pulse. 
The time-dependent wavefunctions are discretized on a spatial grid consisting 
of $n_\text{grid}=1024$ 
equidistant points in the interval $[-L,L] = [-150,150]$ 
for the Coulomb model and $[-20,20]$ for the Morse model,
in both spatial directions.
The ground state wave functions are determined by minimizing the 
expectation values of the respective field-free Hamiltonians using the conjugate gradient method with 
inverse iterations.
In case of the Coulomb system this results in a 2D $1s$ orbital, while
for the Morse potential it yields an annular-shaped wave function with a peak at $r_e$,
i.e. at the ground-state equilibrium internuclear distance.
The further time-propagation is performed using the second-order 
split-operator approach with the kinetic energy operator approximated using Fast Fourier Transform.
The adopted time step is $h = 0.01$ a.u. for the Coulomb model and 
$h = 0.05$ a.u. for the Morse model.
We consider the dynamics for times $0 \leq t \leq 100$ a.u. for the Coulomb model and $0 \leq t \leq 
300$ a.u. for the Morse model, including periods of free evolution after the laser pulses.
Some snapshots of the time evolution of the wave
functions for the Coulomb and Morse 
simulations featured in this work can be found 
in Ref.{\citenum{Wozniak20243659}}. 
In both cases the respective wave functions become
increasingly more complicated with many features, 
more deformed and oscillatory, and more diffused as the time-simulation progresses.

\subsection{Fully flexible ECG basis set}

As elaborated in our previous work\cite{Wozniak20243659}, 
complex ECGs with optimizable nonlinear parameters are 
the building blocks of the time-dependent wave function in our approach.
The Adamowicz group has used various types of ECGs to perform 
very accurate calculations of stationary atomic and molecular bound states for over two decades 
\cite{Bubin2013,Kozlowski19932007,Mitroy2013,Bubin2005377}.
A brief review of these different kinds of basis sets
can be found in our previous work \cite{Mitroy2013}.
Here, we employ the same form of the 2D fully flexible ECGs (FFECGs) as in our previous 
paper{\citenum{Wozniak20243659}}, depending on six real nonlinear parameters $\mathbf{x} = ( q_x, q_y, p_x, p_y, a, b)$, i.e.,
\begin{equation}
    \phi(\mathbf{r};\mathbf{x}) = \exp\left[ -(a + \mathrm{i}b)\|\mathbf{r}-\mathbf{q}\|^2 + \mathrm{i} \mathbf{p}\cdot(\mathbf{r}-\mathbf{q})\right].
\label{eq:FFECG}
\end{equation}
As we demonstrated by fitting the time-dependent wave functions 
of the model systems, such FFECGs are very efficient in reproducing various wave
function shapes, including highly complicated, oscillatory ones.
On the other hand, they can also easily approximate the purely real-valued ground state 
wave functions, which makes them a good starting point for the time-propagation.
In particular, the ground state of the Coulomb system centered at $x=y=0$ consists solely of simple 
spherical Gaussians with $a \ne 0$ and $b = p_x = p_y = q_x = q_y = 0$.
For the Morse-model ground state, the FFECGs with $a \ne 0$, $b \ne 0$, and $p_x = p_y = q_x = q_y = 
0$ need to be employed to reproduce the annular shape of its wave function.
However, for the function to remain real, for each value of $a$ two FFECGs with $b$ and $-b$ are used, 
so that their linear combination results in a Gaussian function multiplied 
by a $\sin$ (or $\cos$) function:
\begin{align}
\frac{\phi(-b) - \phi(+b)}{2\mathrm{i}}
= \exp\left\{ -a (x^2 + y^2) \sin \left[b (x^2 + y^2) \right] \right \}
= \exp(-ar^2) \sin(b r^2).
\label{sin}
\end{align}
As function (\ref{sin}) is zero at the center of the coordinate system,
it helps to properly represent the behavior of the Morse ground-state wave function at this point.
Thus, in the expanding the ground-state wave function for the Morse model, 
which a torus shape,
the FFECGs with $a \ne 0$, $b \ne 0$, and $p_x = p_y = q_x = q_y = 0$
are used. 

\subsection{The Rothe variational method for solving the TDSE}

The spatial part of the wave function of the system, $\Psi$, is expanded in terms of the 
symmetry-adapted FFECG basis
functions, $\phi_i({\bf x}_i)$. Each function depends on a set of  
non-linear parameters $\mathbf{x}_i$. These parameters are the matrix elements 
of $\mathbf{A}$ and $\mathbf{B}$ 
and the elements of the $\mathbf{p}$ and $\mathbf{q}$ vectors.
Let us write the spatial part of the wave function as a linear combination
of the FFECGs, suppressing the spatial dependence for simplicity:
\begin{equation}
\Psi(\mathbf{C},\mathbf{x}) = \sum_i \phi_i({\bf x}_i) c_i 
\equiv \phi(\mathbf{x}) \mathbf{C},
\end{equation}
where $\mathbf{C} = \{c_i\}$ is the vector of linear expansion coefficients, and where $\mathbf{x} = \{\mathbf{x}_i\}$ is the collection of all nonlinear parameters, $\mathbf{x}_i$ being the parameters for the function $\phi_i$. We arrange the $\phi_i$ in a row ``matrix'' $\phi(\mathbf{x})$. 
Using the implicit midpoint method with time step $\Delta t$ to discretize the TDSE \emph{in time only} gives a space-continuous Crank--Nicolson scheme,
\begin{equation}
    {\rm i}  \Psi^{n+1} -{\rm i} \Psi^n -
\frac{1}{2} ( {\hat H}(t +\tfrac{1}{2}\Delta t) \Psi^{n+1} + {\hat H}(t + \tfrac{1}{2}\Delta t) \Psi^n)\Delta t = 0.
\end{equation}
where $\Psi^n$ is the time-discrete wavefunction at time $t = n\Delta t$. Given any square integrable $\Psi^n$, the next time-step $\Psi^{n+1}$ exists, since the solution operator is a unitary operator defined everywhere in Hilbert space.

The Rothe method for propagating the TDSE now posits an ansatz $\Psi^{n} = \Psi(\mathbf{C}^{n},\mathbf{x}^{n})$ for all time steps $n$, and variationally optimizes the residual in the semidiscrete TDSE to within a selected error threshold $\epsilon$. We obtain the following optimization problem: Find $(\mathbf{x}^{n+1},\mathbf{C}^{n+1})$ such that
\begin{equation}
\|A \Psi(\mathbf{x}^{n+1},\mathbf{C}^{n+1}) -
  A^\dag \Psi(\mathbf{x}^n,\mathbf{C}^n) \|  < \epsilon. 
\end{equation}
Here, $A = I + \frac{{\rm i} \Delta t}{2} {\hat H}(t+\Delta t/2) $. If the numerical optimization is not able to meet the desired tolerance $\epsilon$, the length of the FFECG expansion is increased, allowing more details in the wavefunction to be resolved and thus lowering the residual. We return to the precise specification of this procedure below.

Since $\mathbf{C}^{n+1}$ appears in a linear manner, we can use the standard normal equation approach to solve for these, given a current guess for $\mathbf{x}^{n+1}$. This is referred to as the \emph{variable projection method}\cite{golubSeparableNonlinearLeast2003}.
The solution reads
\begin{equation}
    {\mathbf{C}}^{n+1}  = {\bf S}_A^{-1} {\mathbf \phi}^{\dagger} A^\dag \Psi^{n},
\end{equation}
where a weighted overlap matrix has been introduced, with elements
\begin{equation}
(S_A)_{ij} = \langle{\phi_i|A^\dag A |\phi_j}\rangle.
\end{equation}
Furthermore, for any wavefunction $\Phi$, we have $\phi^\dag\Phi = \mathbf{b}$, with $b_i = \langle \phi_i|\Phi\rangle$, a vector of length equal to the number of FFECG basis functions.
We introduce an orthogonal projection operator given as
\begin{equation}
P_A = (A {\mathbf \phi}) {\bf S}_A^{-1} (A {\mathbf \phi})^{\dagger} = \sum_{ij} A |\phi_i\rangle (S_A^{-1})_{ij} \langle \phi_j | A^\dagger.
\end{equation}
The optimization problem now reduces to: Find $\mathbf{x}^{n+1}$ such that
\begin{equation}
F({\bf x}^{n+1}) \equiv ||(P_A({\bf x}^{n+1}) - I)A \Psi^n||  < \epsilon. \label{eq:Rothe functional}
\end{equation}
We now use the Gauss--Newton method to minimize the Rothe cost function $F$. The minimization is carried out as long as it takes to
lower the value of $F({\bf x}^{n+1})$ below the threshold, increasing the expansion length as needed. In the result section,
different values of $\epsilon$ are used in the time-propagation to test how the
accuracy of the results (as compared with the results of the grid calculation) depends 
on the value of the threshold.

\subsection{Computational details}

The calculations involving wave-packet time propagation 
is performed for two 2D models mentioned before.
In 2D, the FFECGs are given by Eq.~\eqref{eq:FFECG}. Each 2D FFECG 
depends on six real non-linear parameters,
$a$, $b$, $p_x$, $p_y$, $q_x$, and $q_y$.
As the time-propagation calculation for both models are
initiated with the corresponding ground-state
wave functions, which are real and spherically 
symmetric, appropriate FFECGs need to be used
in the wave-function expansion. 
For the Coulomb model, the FFECGs
are simple spherical Gaussians centered at $x=y=0$
with $a \ne 0$ and $b = p_x = p_y = q_x = q_y = 0$. 
A linear combination of FFECGs
of this type should provide a very accurate representation of 
the ground-state wave function.
The  linear and non-linear parameters of FFECGs for such a representation
are obtained in the present work with the use of the variational method.
The minimization of the total energy of the system with respect to the
linear and non-linear FFECGs parameters results is a compact 
wave function that is used as a starting point for the time-propagation.

The minimization of the Rothe functional that involves optimization of all
linear and nonlinear parameters of the enlarged 
FFECG basis set continues for some number 
of time steps until it is determined
that the optimization process is unsuccessful because 
the accuracy threshold cannot be met.
At that point new FFECGs are added to the basis set
and the optimization continues.
The results for the Coulomb and Morse models
are shown in Fig.~\ref{fig:cost}.
In the figures, the results obtained in the
grid calculations are compared with the corresponding results obtained 
with the Rothe time-propagation method.
The results obtained for the $\epsilon$ threshold values of $10^{-3}$, $10^{-4}$, and $10^{-5}$
are shown. As expected, achieving the lowest 
threshold required a large FFECG basis set (the green cureve).
For the highest threshold the number of Gaussians for 
the two models remains the same as used to calculate
the starting ground-state wave function. The plots of the cost functions show that, as expected, 
the lowest value of the threshold allows the Rothe method to 
produce virtually exact solution of the TDSE.
With the less tight threshold the cost function is visibly 
less accurate. The same trend is also showing
in the plots of the error.

In general, the Morse model seems to be somewhat 
more difficult to describe using FFECGs and the Rothe method
than the Coulomb model. 
The reason for this behavior can be attributed to the maximum 
of the density of the second nucleus in the Morse model being 
shifted away from the reference nucleus by some distance 
(in the ground-state this distance is approximately equal to the equilibrium internuclear 
distance). This type of shifting does not happen in the Coulomb model. 
The shifting of the density maximum away from the reference nucleus required 
including more ECGs in the wave function. 
However, if this is done, the ECG expansion of the 
wave packet in the Morse model should 
be equally accurate as it is for the Coulomb model.

The accuracy of the results obtained from the Rothe-propagation calculations for the
Coulomb and Morse models can also be verified
by calculating some other properties than the energy 
using the wave packets obtained for some selected
time steps and compare of their time-dependent dynamics
with the dynamics obtained using the grid method. 
The properties calculated in this work include the following:
These other observables calculated in this work include the expectation 
value of the field-free Hamiltonian, 
the ground-state survival probability (i.e., the square of
the autocorrelation function),
the dipole moment expectation value, 
and the expectation value of the squared 
$z$-component of the angular momentum. Thus, we evaluate
$\langle \Psi (t) | \hat H_0 | \Psi(t)\rangle$,
$\langle \Psi (0) | \Psi(t)\rangle$,
$\langle \Psi (t) | \hat x | \Psi(t)\rangle$, and
$\langle \Psi (t) | \hat L^2_z | \Psi(t)\rangle$ during time evolution.
Plots showing the results of the property calculations 
for the two models are shown in Fig.~\ref{fig:obser}.
The properties are calculated for the wave packets obtained
from the Rothe propagation carried out with the three
values of the accuracy threshold, $\epsilon$, mentioned above in the
discussion concerning Fig.~\ref{fig:cost}. Besides the property values plotted
as a function of time, plots of the difference of the property values obtained
from the Rothe propagations and the values obtained from the reference grid calculations
are also shown in Fig.~\ref{fig:obser}. The plots of the differences clearly show that
tightening the convergence of the optimization of the FFECG non-linear parameters
in the Rothe calculations
by lowering the value of the $\epsilon$ renders the values of the properties
that are in increasingly better agreement with the grid results.
As one can see in Fig.~\ref{fig:obser}, the smallest 
$\epsilon$ renders for both models the property results
virtually the same as the grid results. The results obtained with the wave packets 
generated with the Rothe method with two larger $\epsilon$ values are visibly not as good 
as the results obtained for the smallest $\epsilon$.

Lastly, the FFECG wave packets obtained in the Rothe propagation at two selected moments of time
for the Coulomb and Morse models are compared with wave packets obtained in the propagation on the 
grid. The comparison is shown for the two models in 
Figs.~\ref{fig:wave_packet_Coulomb} and \ref{fig:wave_packet_Coulomb}.
In each figure there are three columns of frames. The first three correspond to the three 
accuracy thresholds used in the Rothe propagation. The last column of frames shows
the wave packet obtained from the grid simulations. 
The purpose of showing the wave packet is to again demonstrate the agreement between the
results obtained in the Rothe time propagatin and the reference grid propagation.
As one can see, as the accuracy Rothe threshold decreases and the minimalization of the 
Rothe funcional becomes tighter the wave packets for both models more accurately resemble
the wave packets obtained from the grid calculations.
The figures also show how significant is the deformation of the ground state wave function
for each of the models when it interacts with the laser pulses used in the 
calculations.

The first steps in the FFECG calculations involves 
determining the nonlinear parameters that define the ground state 
wave functions of the two model systems, as described in the earlier subsection.
In our previous work, this was done by fitting a predefined number 
of Gaussians to the grid-based ground states.
However here we follow a different approach, namely for 
both Morse and Coulomb system we explicitly 
optimize the ground-state Hamiltonian eigenvalue:
\begin{equation}
||{\hat H_0}{\Psi}(\mathbf{x}^0, \mathbf{C}^0) - E_0 \Psi(\mathbf{x}^0, \mathbf{C}^0)|| = \min !
\end{equation}
The minimization is carried out with respect to both linear and nonlinear parameters 
($a$ in case of the Coulomb model and $|b|$ in case of the Morse model).
There are two reasons behind this.
Firstly, by doing so, we decouple the Rothe simulations 
from the grid-based simulations, allowing the grid and 
Gaussian wave functions to evolve completely independently of each other.
Secondly, due to numerical differences between the grid 
and Gaussian representation of the wave 
function, the ECG ground state obtained by fitting may 
differ slightly from the state with optimal $E_0$.
Since the Rothe method relies heavily on the minimization of 
the Hamiltonian value, any deviation from the actual ground 
state energy value may then be propagated during the time evolution.
For the Coulomb model we use six FFECGs for the ground 
state and arrive at $E_0 = -0.6554858$~Ha vs the numerical value of $-0.6554864$ Ha
obtained from the grid reference calaculation. For the Morse model we use eight FFECGs, 
obtaining $E_0 = -0.1639631$~Ha vs the numerical value of $-0.1639638$~Ha.

The minimization of the Rothe cost function \eqref{eq:Rothe functional} with respect of
the non-linear parameters of FFECGs at every time step is then performed 
with the standard least-squares method implemented in
our in-house code that employs the SciPy Python library\cite{SciPy}.
The matrix elements of the Hamiltonian, Hamiltonian square, and overlap 
over the FFECGs are calculated by first transforming the Gaussians
to functions defined on the 2D grid and then carrying out the integration
numerically. This is a similar approach as the one used before 
in a development involving slater orbitals and grid-based 
molecular orbitals for diatomic systems \cite{Adamowicz1984373}.
At the starting point, the FFECGs are the same as used to expand the ground-state
wave function but the parameters that are set to zero for the 
ground-state wave function are now unfrozen and become optimized.
At each time step the current value of the cost function is monitored.
If at some point it exceeds a predefined threshold value, 
new Gaussian functions are added to the basis set and the total wave function is reoptimized.
This procedure is repeated until the cost function decreases below the threshold value.
The choice of the initial parameters of the added basis functions is 
subjective, but we applied a following scheme, which we found to be quite effective.
For the Coulomb model simulations we add a single FFECG at a time, with the nonlinear parameters 
(denoted with $'$) being the averages of the respective nonlinear parameter values already present in 
the basis set:
\begin{equation}
\phi\left(a'=\frac{\sum_i c_i a_i}{\sum_i c_i}, b'=\frac{\sum_i c_i b_i}{\sum_i c_i}, p_x = \frac{\sum_i c_i p_{x,i}}{\sum_i c_i}, p_y = \frac{\sum_i c_i p_{y,i}}{\sum_i c_i}, q_x = \frac{\sum_i c_i q_{x,i}}{\sum_i c_i}, q_y = \frac{\sum_i c_i q_{y,i}}{\sum_i c_i}\right).
\end{equation}
For the Morse model we use a similar approach, 
except that we add two FFECGs at a time, with $b'$s being the 
weighted averages of either positive or negative values of $b$ in the basis set:
\begin{align}
&\phi\left(a'=\frac{\sum_i c_i a_i}{\sum_i c_i}, b'=\frac{\sum_{i,b_i>0} c_i b_i}{\sum_{i,b_i>0} c_i}, p_x = \frac{\sum_i c_i p_{x,i}}{\sum_i c_i}, p_y = \frac{\sum_i c_i p_{y,i}}{\sum_i c_i}, q_x = \frac{\sum_i c_i q_{x,i}}{\sum_i c_i}, q_y = \frac{\sum_i c_i q_{y,i}}{\sum_i c_i}\right), \nonumber \\
&\phi\left(a'=\frac{\sum_i c_i a_i}{\sum_i c_i}, b'=\frac{\sum_{i,b_i<0} c_i b_i}{\sum_{i,b_i<0} c_i}, p_x = \frac{\sum_i c_i p_{x,i}}{\sum_i c_i}, p_y = \frac{\sum_i c_i p_{y,i}}{\sum_i c_i}, q_x = \frac{\sum_i c_i q_{x,i}}{\sum_i c_i}, q_y = \frac{\sum_i c_i q_{y,i}}{\sum_i c_i}\right).
\end{align}
This is to ensure that the added functions reflect the (possibly distorted) 
annular shape of the rovibrational wave function.
For both models we test three cost-function 
threshold values: $10^{-3}$, $10^{-4}$, and $10^{-5}$, 
for each value we monitor the value of the cost function, 
the error value (measured as the integral of the difference between the 
reference grid wave function and the Gaussian wave function), 
and the number of FFECGs in the basis set.

Initially, we attempted to conduct the Rothe propagations 
using the same time steps as in the grid-based propagations.
However, we found the simulations to be numerically unstable, 
with the minimization procedure failing when the external field reached larger values. Consequently, the time steps are reduced to $0.002$ a.u. 
for the Coulomb model and to $0.01$ a.u. for the Morse model.

\section{Results}

The results of the simulations performed in this work are presented 
in Figs. \ref{fig:obser}-\ref{fig:cost}.
In Fig.~\ref{fig:obser}, the trajectories of four time-resolved observables are shown. These are 
the expectation value of the ground-state Hamiltonian, 
the projection of the time-dependent wave function on the respective ground state, 
the dipole-moment expectation value, 
and expectation value of the squared angular-momentum projection onto 
the axis perpendicular to the simulation plane.
Fig.~\ref{fig:wave_packet_Morse} displays arbitrarily selected 
snapshots from the wave function trajectories of the Coulomb and Morse 
models obtained from grid-based propagation and the Rothe propagation 
with three different cost-function thresholds.
Finally, in Fig.~\ref{fig:cost}, the time-resolved cost function, the wave function error, 
and basis set size are shown.

First and foremost, the simulations with the tightest adopted 
cost threshold $\epsilon$ for both models yield time-resolved observables that are virtually 
indistinguishable from those calculated using the grid-based approach.
This is particularly evident in the plots of the observable errors 
(Fig.~\ref{fig:obser}), all of which consistently maintain a negligible 
level throughout the entire propagation periods.
This consistency is further confirmed by comparing 
the wave function shapes in the two rightmost columns of  Fig.~\ref{fig:wave_packet_Morse}.
This result is the main outcome of this work, as it directly 
confirms that the Rothe propagation algorithm — and, by extension, the reformulation 
of the TDSE as an optimization problem — is fundamentally equivalent 
to traditional time-discretization approaches which can all be seen as based on the Magnus formalism.
Furthermore, it reinforces the conclusions of our previous work, 
that the FFECG basis set can perform on par with spatial grids.
These results are even more striking when considering that 
the FFECG propagations and the grid-based propagations 
are performed completely independently, without any exchange of 
information about the wave-function shape neither during the ground-state 
computation nor at any point during the propagation.
However, as expected, this excellent agreement comes at a cost: 
a tighter cost threshold leads to a rapid increase of the basis set size.
By the end of the simulations with threshold $10^{-5}$, 
the number of FFECGs exceeds 50 for the Coulomb model 
and 80 for the Morse model, impacting the CPU-time use per time step.

Loosening the threshold to $10^{-4}$ significantly reduces 
the final number of Gaussians needed -- approximately 15 for the 
Coulomb model and 55 for the Morse model -- while still producing results 
very close to the grid-based ones in terms of both the observables and the 
wave functions.
However, some regions of lesser agreement with the grid results
can be observed in Fig.~\ref{fig:wave_packet_Morse}.
For instance, in the snapshots from the Coulomb model trajectory, 
this simulation variant fails to reproduce the wave-function maximum 
near the nucleus when the wave function becomes highly dissipated.

Finally, the results for the simulations with threshold $10^{-3}$ 
are best described as qualitatively or semi-quantitatively correct 
rather than quantitatively correct.
Noticeable deviations from the grid curves are apparent 
for the observables in Fig.~\ref{fig:obser}, and for the wave-function shapes in 
Fig.~\ref{fig:wave_packet_Morse}. The Rothe results only roughly approximate the grid results.
Nevertheless, the overall shapes of the curves are preserved in most observables, 
with only two exceptions, the $\langle x \rangle$ and $\langle L_z^2 \rangle$ 
trajectories for the Morse model.
The discrepancy in $\langle x \rangle$ in the Morse model can be attributed to 
the nature of the dissociation.
Due to the high nuclear masses, even when the molecule is excited 
to very high energies, as occurs in the simulation, the majority of 
the wave function remains well-localized, while only a small portion undergoes scattering.
Describing both components simultaneously using a small number of Gaussians 
proves challenging, as seen in the leftmost column of Fig.~\ref{fig:wave_packet_Morse}.
This leads to the dipole moment stabilizing at a smaller value than in the reference simulation.
The insufficient number of basis functions also accounts 
for the disagreement in the angular-momentum-projection expectation value.
In the calculations, we use only $S$-type FFECGs 
(not multiplied by spherical harmonics), meaning that the angular component 
of the wave function arises solely from mixing between the 
plane-wave components of the Gaussians.
If the basis set contains too few frequencies ($q_x$ and $q_y$ values), 
the resulting wave function may struggle to incorporate higher angular momenta.
This is evident in Fig.~\ref{fig:obser}, where $\langle L_z^2 \rangle$ starts 
to rapidly decrease immediately after the laser pulse ends.
This suggests that the wave function undergoes a reflection from the basis set 
boundary, similar to effects observed in HHG calculations with Gaussian basis sets 
\cite{Coccia2016,wozniak2021,wozniak2022,wozniak2023}.
On the other hand, the trajectories of $\langle H_0 \rangle$ 
and $\langle \Psi_0 | \Psi(t) \rangle$ in the Morse model, as well as 
the trajectories of all four observables in the Coulomb model, 
remain within a few percent error from the reference grid curves.
Notably, for both models, the threshold of $10^{-3}$ did not necessitate 
any expansion of the basis set, i.e. the entire propagation is conducted 
using only the initial six Gaussians for the Coulomb model and eight Gaussians for the Morse model.
Of course, in every set of the propagations, the non-linear parameters of the Gaussians are reoptimized. 
As a result, the propagations took only a few hours to complete.
Given such low computational complexity, the observed shortcomings may 
be considered acceptable trade-offs.

Overall, the basis expansion rate (the average number of the added 
functions per time step) is noticeably higher for the Morse model than for the Coulomb model.
This indicates that the ro-vibrational dynamics of the Morse model is 
inherently more difficult to describe using Gaussians.
This observation aligns with our previous findings when 
fitting the numerical time-dependent wave functions with FFECGs.
This can be attributed to the aforementioned necessity of 
simultaneously describing both the well-localized, annular-shaped part 
of the wave function and the highly diffuse scattering part.
Additionally, the maximum density of the second nucleus 
in the Morse model is shifted away from the reference nucleus by a certain 
distance, making it more challenging to represent using single-centered Gaussians.
Another factor of significance is that, in the Morse simulation, 
two Gaussians are added at a time. 
However, when these results are translated to real time, 
the difference diminishes or even shifts in favor of the Morse model.
For instance, at $t = 45$ a.u., with the threshold of $10^{-4}$ 
the wave functions consist of 15 and 14 FFECGs for the Coulomb and Morse models, respectively.
For the same time, with the threshold of $10^{-5}$, the wave functions 
consist of 48 and 32 FFECGs for the Coulomb and Morse models, respectively.
This suggests that the more complex nature of the nuclear dynamics 
is somewhat offset by the slower motion of the nuclei due to their higher masses.
While this is only a rough estimation -- 
given the significantly different shapes of the laser pulses applied to 
both models and the dynamics they induce -- it may be seen as a promising 
indication for future simulations involving both the nuclear and 
electronic motions, as it implies that the basis of the electronic and 
nuclear wave functions should be augmented at a similar rate.

\begin{figure}
  \centering
  \subfloat{\includegraphics[width=0.85\textwidth]{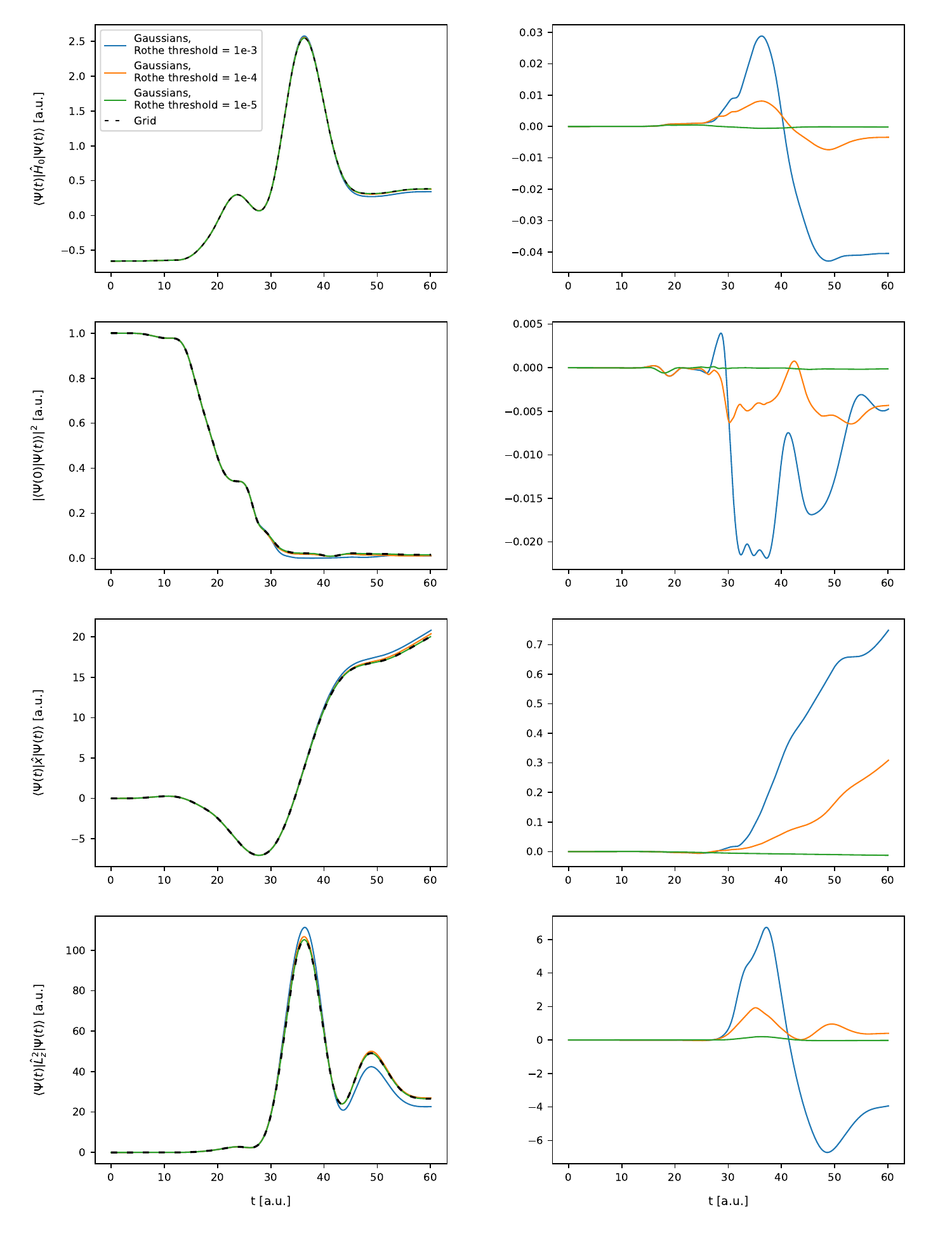}}
  \captionsetup{labelformat=empty}
  \caption{}
\end{figure}  
\begin{figure}
  \centering  
  \subfloat{\includegraphics[width=0.85\textwidth]{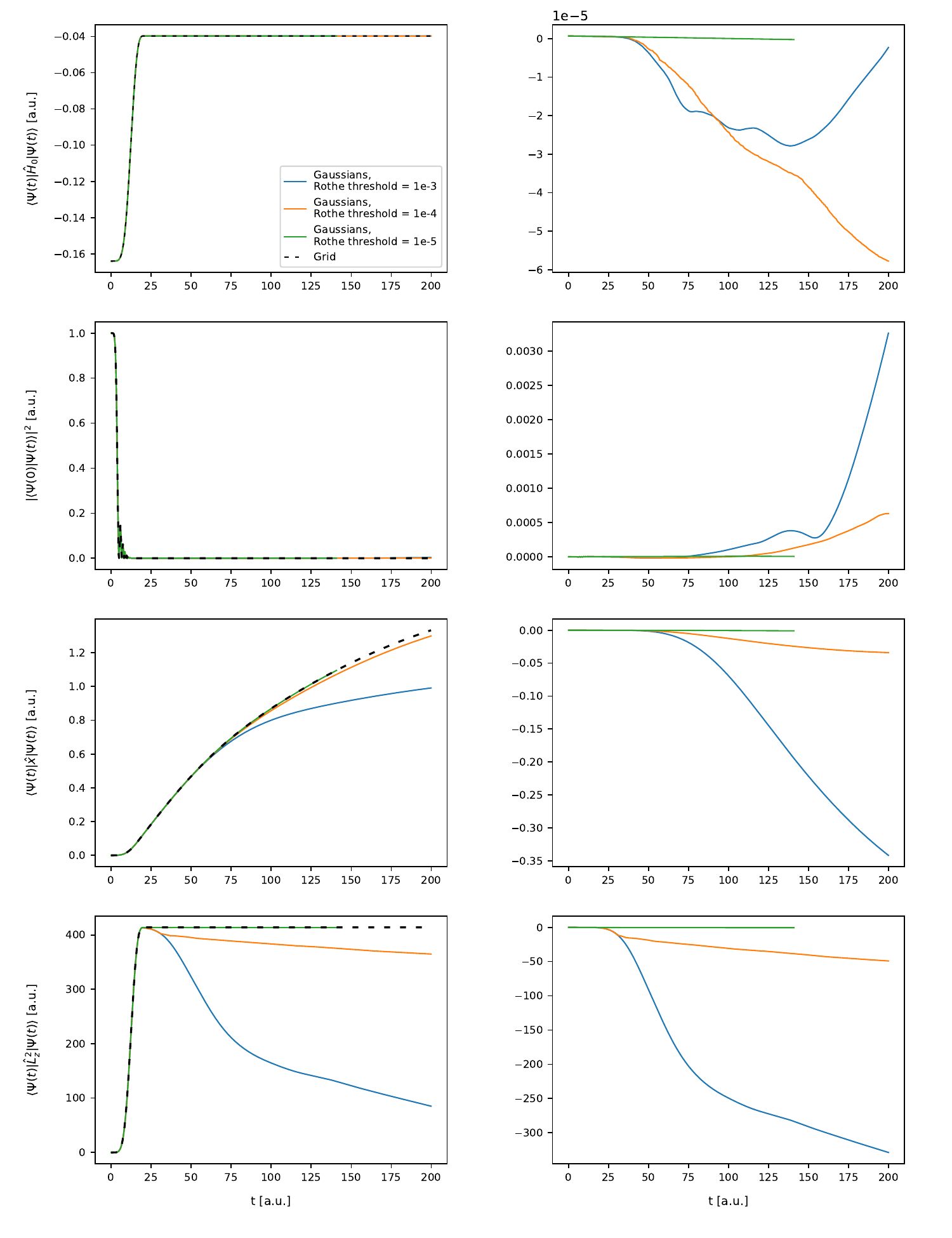}}
  \captionsetup{labelformat=empty}
  \caption{}
\end{figure}
\begin{figure}
  \centering
  \caption{The results of the Rothe time propagation that employs FFECG basis 
  for the Coulomb model (upper eight frames) and for the Morse model (lower eight plots). 
  The plotted results concern the following expectation values: 
  $\langle \Psi (t) | \hat H_0 | \Psi(t)\rangle$,
  $\langle \Psi (0) | \Psi(t)\rangle$,
  $\langle \Psi (t) | \hat x | \Psi(t)\rangle$, and
  $\langle \Psi (t) | \hat L^2_z | \Psi(t)\rangle$.
  The plots on the left for each model show the expectation values as functions of time for three values of $\epsilon$.
  The plots on the right show the error with respect to the exact expectation values obtain using the grid method.}
  \label{fig:obser}
\end{figure}

\begin{figure}
  \centering
  \subfloat{\includegraphics[width=0.85\textwidth]{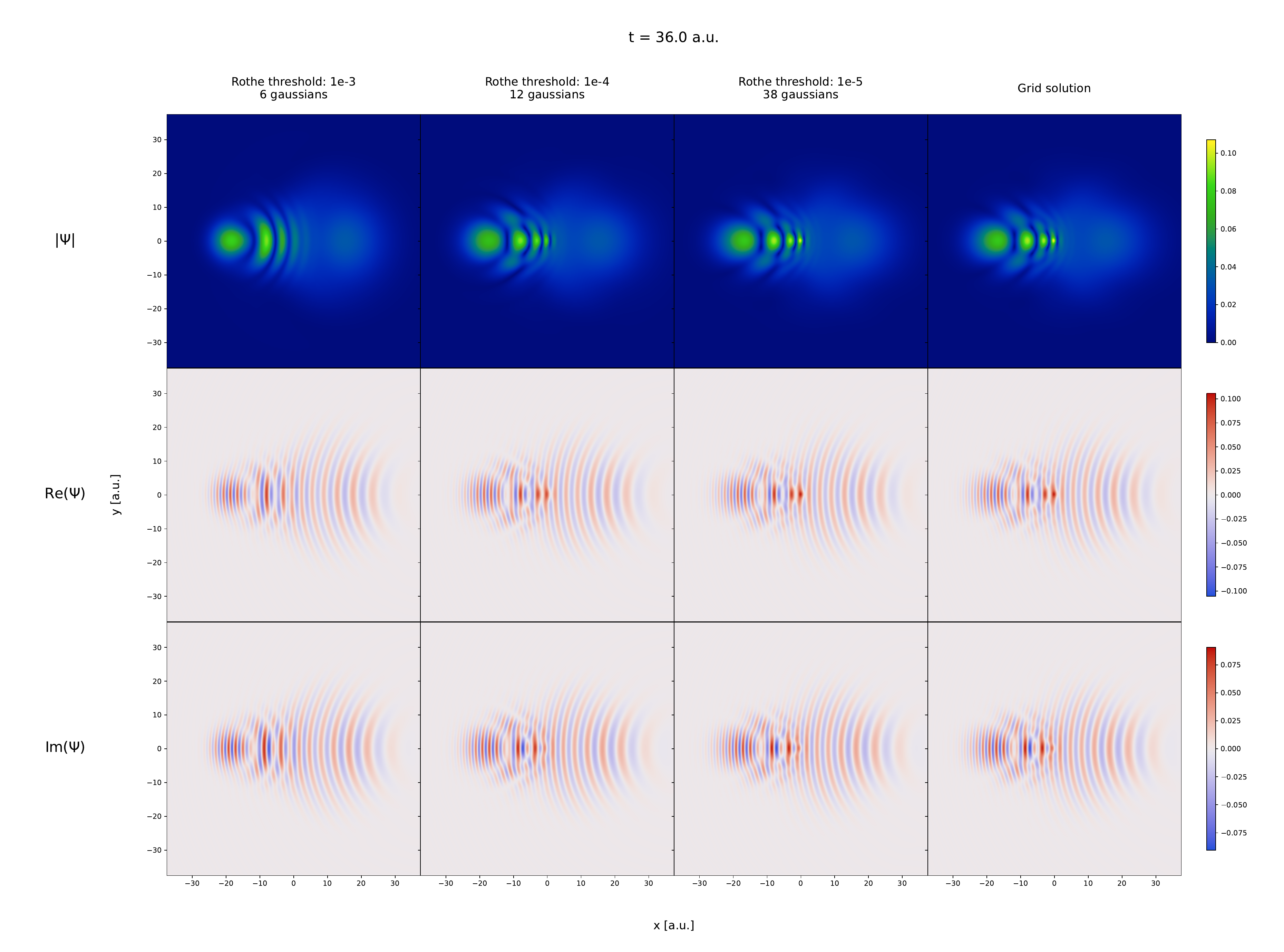}}
  \captionsetup{labelformat=empty}
  \caption{}
\end{figure}
\begin{figure}
  \centering  
  \subfloat{\includegraphics[width=0.85\textwidth]{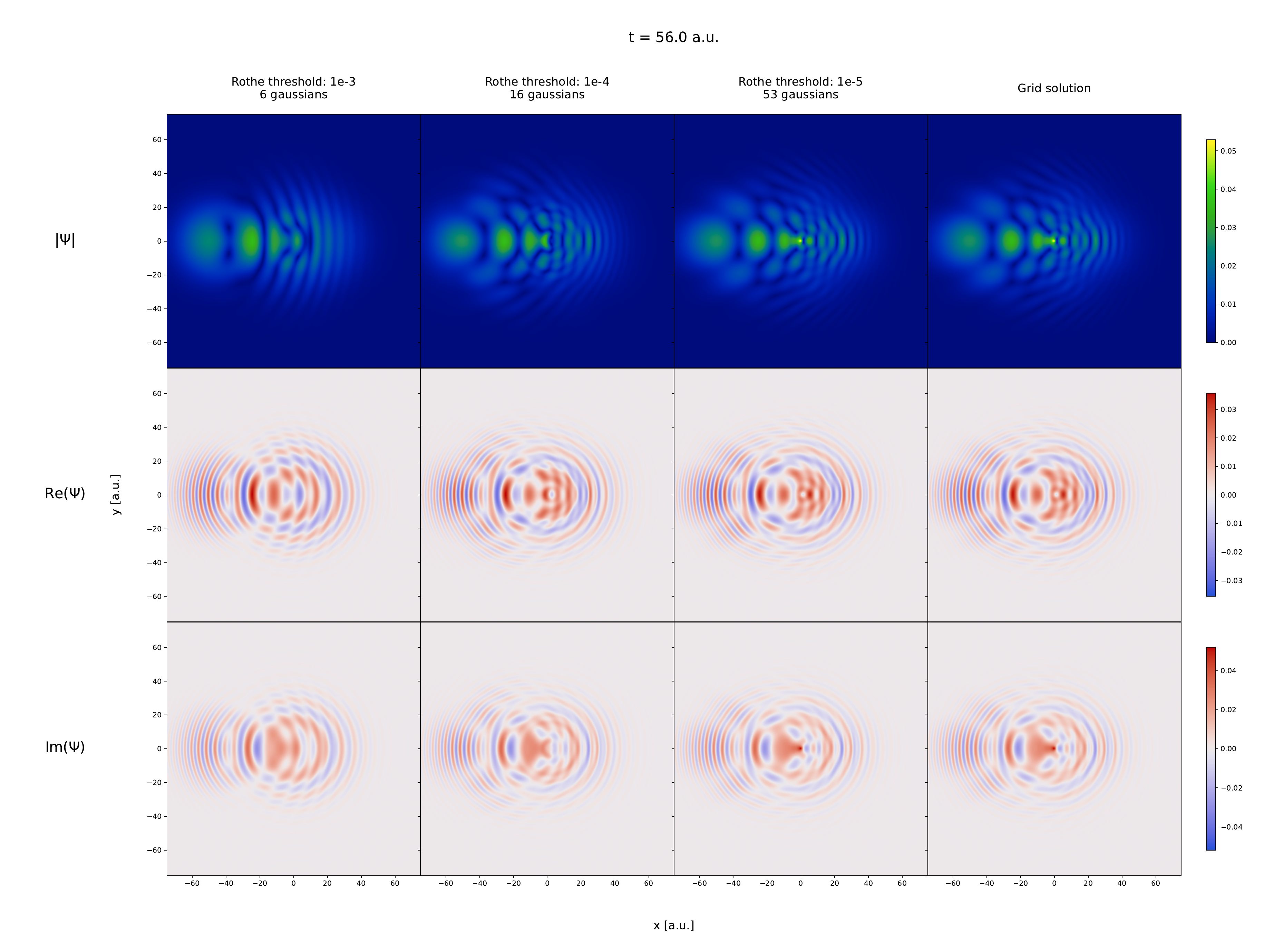}}
  \captionsetup{labelformat=empty}
  \caption{}  
\end{figure}
\begin{figure}
  \centering
  \caption{
  The wave packet, $\Psi$ for the Coulomb model pictured for time equal to 36.0 
  and 56.0 a.u. using plots 
  of $|\Psi|$, ${\rm Re}(\Psi)$, and ${\rm Im}(\Psi)$. 
  The wave packets shown in the first, second, and third columns of 
  frames are obtained in the Rothe propagation using the accuracy  
  threshold values of $10^{-3}$, $10^{-4}$, and $10^{-5}$, respectively.
  The last column of frames shows the wave packet obtained from the grid calculations.}
  \label{fig:wave_packet_Coulomb}
\end{figure}
\begin{figure}
  \centering
  \includegraphics[width=0.99\textwidth]{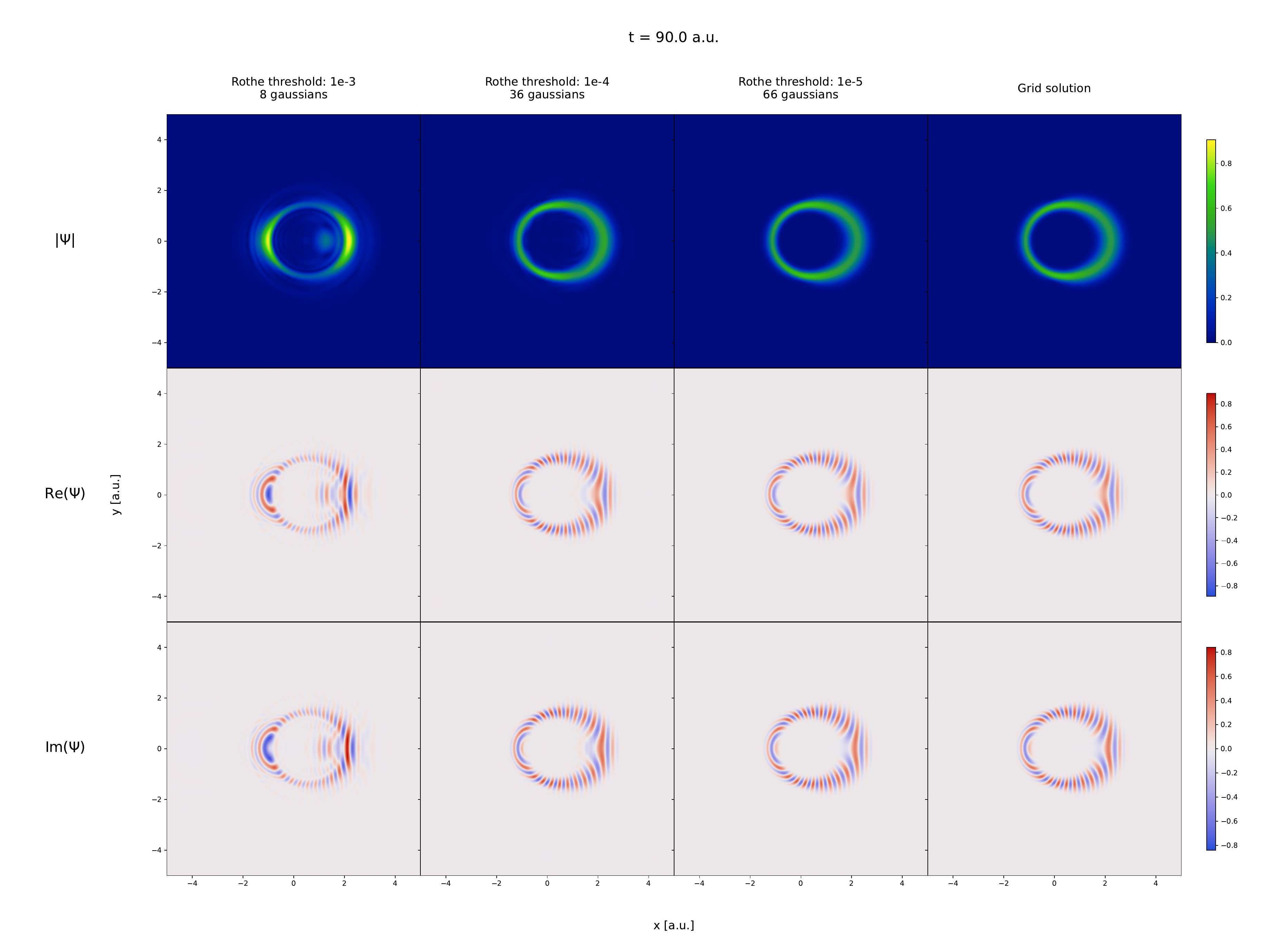}%
    \captionsetup{labelformat=empty}
  \caption{}
\end{figure}
\begin{figure}
  \ContinuedFloat
  \centering  
  \includegraphics[width=0.99\textwidth]{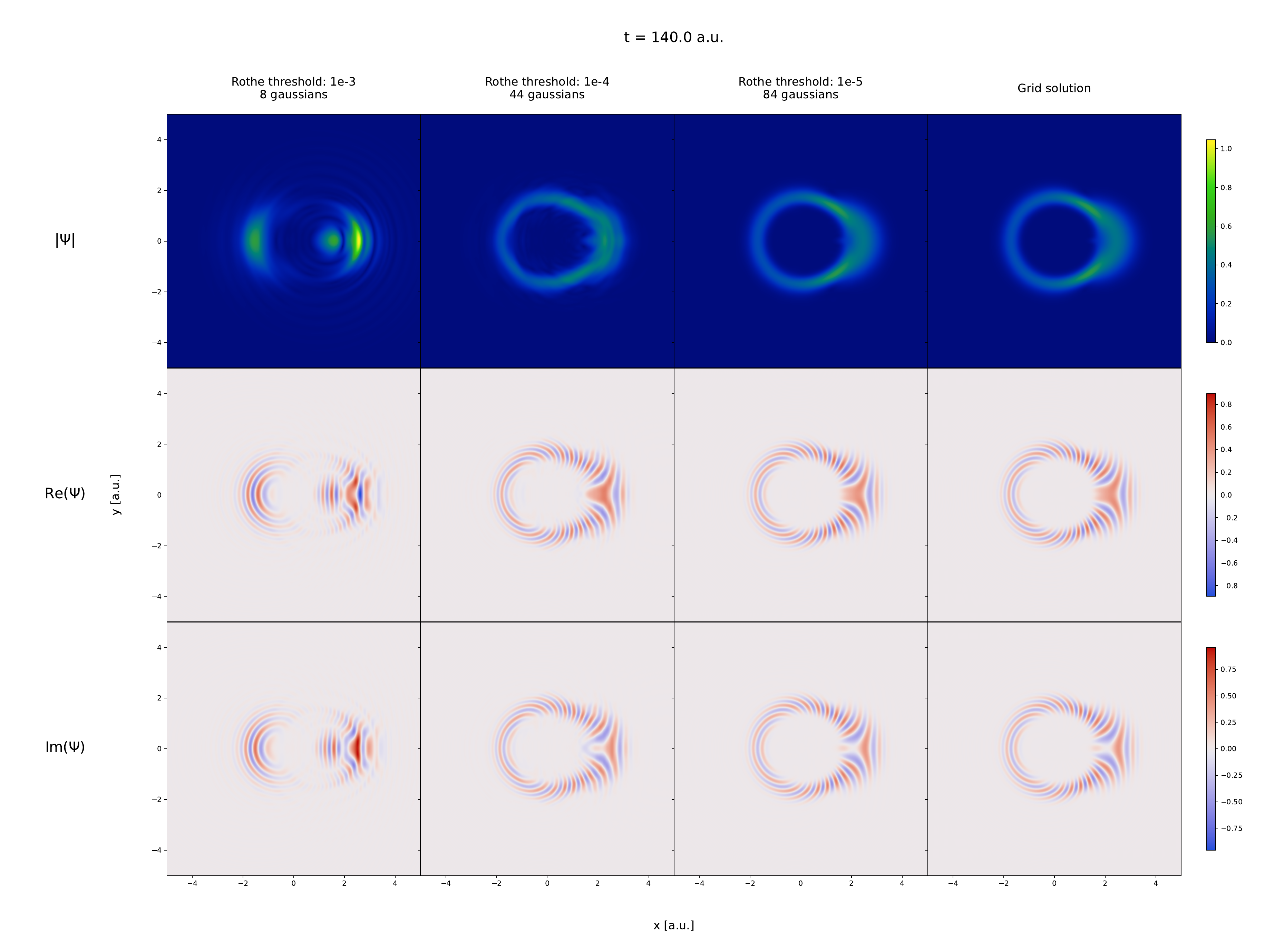}
\end{figure}
\begin{figure}
  \centering
  \caption{
    The wave packet, $\Psi$ for the Morse model pictured for time equal to 90.0 
  and 140.0 a.u. using plots 
  of $|\Psi|$, ${\rm Re}(\Psi)$, and ${\rm Im}(\Psi)$. 
  The wave packets shown in the first, second, and third columns of 
  frames are obtained in the Rothe propagation using the accuracy  
  threshold values of $10^{-3}$, $10^{-4}$, and $10^{-5}$, respectively.
  The last column of frames shows the wave packet obtained from the grid calculations.}
  \label{fig:wave_packet_Morse}
\end{figure}

\begin{figure}
  \centering
  \subfloat{\includegraphics[width=0.85\textwidth]{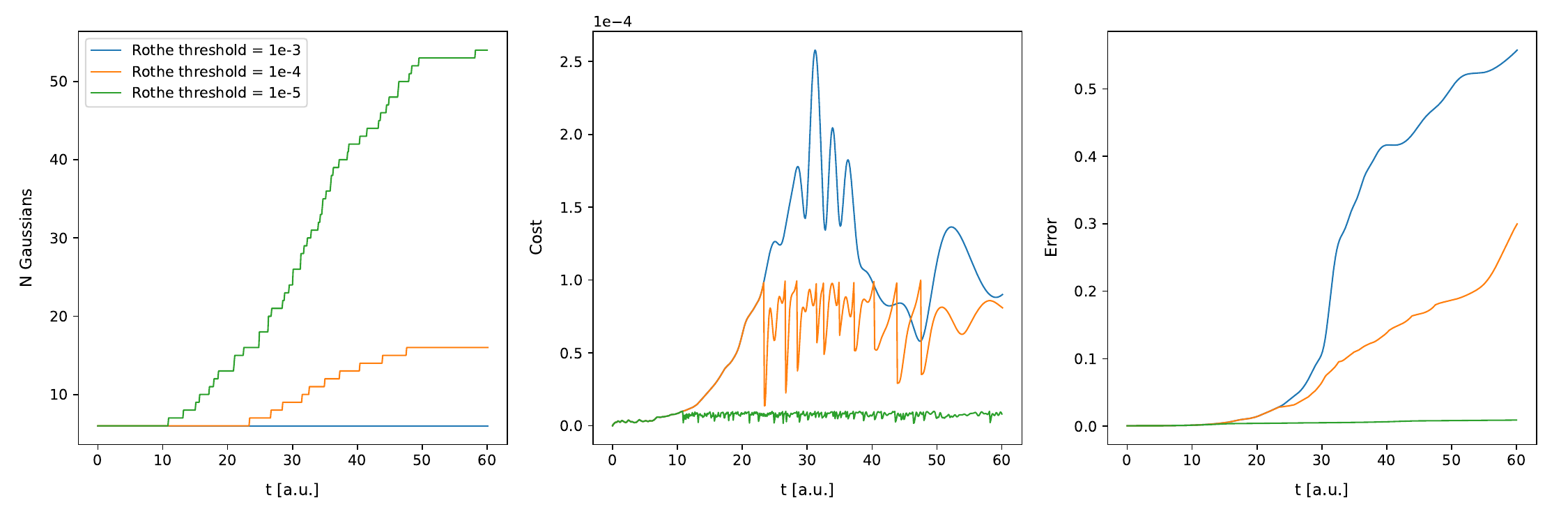}}
  \captionsetup{labelformat=empty}
  \caption{}
  \centering  
  \subfloat{\includegraphics[width=0.85\textwidth]{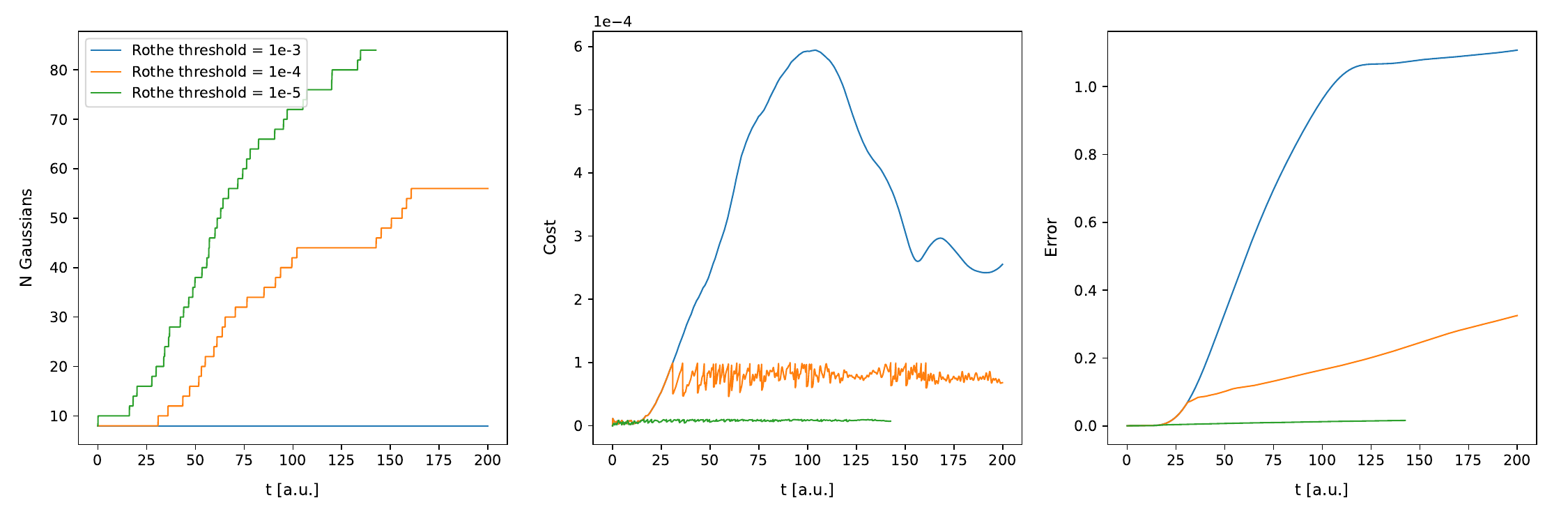}}
  \captionsetup{labelformat=empty}
  \caption{}
\end{figure}
\begin{figure}
  \ContinuedFloat
  \centering
  \caption{ The results of the Rothe time propagation that employs FFECG basis functions for expending the wave packet for the Coulomb model (upper three plots) and for the Morse model (lower three plots). Time is on the horizontal axis in all plots. The first plot in each group shows the number of FFECGs in the basis set that is needed to achieve certain level of accuracy in the calculation as define by the value of the accuracy threshold, $\epsilon$. The second plot is each group shows cost function variation with time for each of the three values of $\epsilon$. The third plot in each group shows the error of the Rothe propagation with respect to the grid propagation.}
  \label{fig:cost}
\end{figure}

\section{Conclusion} \label{conclusion}

The Rothe propagation method is used to obtain solution 
for the time-dependent Schr\"{o}dinger equation
for two 2D model systems that represent features
which appear in non-Born-Oppenheimer quantum-dynamics
time-propagation of the wave packet
representing a diatomic neutral molecule interacting
with a short intense laser pulse.  In non-Born-Oppenheimer 
dynamics, all particles, i.e. the nuclei and electrons are treated
on an equal footing.
In the Rothe propagation, the wave packet representing the system
is expanded in terms of the general 
fully-flexible 2D explicitly correlated Gaussians 
(FFECGs) with complex exponential parameters.
The Rothe time-propagation procedure, also known as the adaptive method 
of time layers\cite{Deuflhard2012_}, is based on a semi-discrete time-dependent Schrödinger equation using the implicit midpoint rule, and variationally minimizing its residual.
The minimization is performed with respect to the linear
expansion coefficients in the linear expansion of 
the wave packet in terms of FFECGs and with respect to the
FFECG non-linear parameter.
The procedure involves growing the basis set of the Gaussians
to provide a uniformly good representation of the time-evolving
wave packet. 
The Rothe results are compared with the results
obtained from the grid time propagation.
Besides comparing the energy values and the wave functions,
other properties are also calculated and used to assess the 
quality of the Rothe results.
The comparison of the results show reliable performance of the
Rothe method in solving the time-dependent Schr\"{o}dinger equation 
for both model systems considered in this work.

Finally, this work represents a preliminary step in the application 
of FFECGs to describe the coupled electronic-nuclear dynamics 
in atomic and molecular systems.
In the future work involving FFECGs and the non-BO 
nuclear-electronic quantum dynamics, the Rothe method 
will be employed to propagate the wave packet representing 
a real chemical system. 
The Rothe approach 
is a viable alternative to the standard real-time propagation 
techniques based on the McLachlan or time-dependent variational principles that involve numerically solving very stiff ordinary-differential equations. 
The FFECG optimization protocol developed in this work to
 minimize the Rothe functional with respect to the linear 
 and non-linear parameters of the FFECGs
 will be applied in our future work to time propagate
 FFECG wave packet representing real atomic and molecular systems.
 In the propagation, we will employ analytical gradients of the Rothe functional
 determined with respect to the FFECG non-linear parameters,
 to aid the optimization. The approach involving the analytical energy gradient
 in variational ECG calculations of ground and excited stationary states
 has been successfully implemented in calculations for small atoms and
 molecules. 
The use of the gradient has significantly expedited the 
functional minimization and enabled to obtain non-BO energies 
and the corresponding wave functions whose accuracy by far exceeds 
the results obtained by other techniques. 
Implementation of the gradient in the Rothe method will require 
derivation of new algorithms to calculate gradient matrix elements. 
As in the case of the variational algorithm in the calculations of stationary 
states, this can be done using the matrix differential 
calculus. An effective implementation of the Rothe method 
to real chemical systems will also require 
highly efficient parallelization and vectorization, 
as well as implementation of GPU-enabled strategies.

\begin{acknowledgement}
This work was supported by the Research Council of Norway through 
its Centres of Excellence scheme, project no. 262695.
Support from the National Science Foundation (grant no. 1856702) is also acknowledged.
The calculations presented in this work were carried out 
using resources provided by University of Arizona Research 
Computing and by Wroclaw Centre for 
Networking and Supercomputing, Grant No. 567.
\end{acknowledgement}

\bibliography{a}

\end{document}